# VaultFS: Write-once Software Support at the File System Level Against Ransomware Attacks

Pasquale Caporaso, Giuseppe Bianchi and Francesco Quaglia

**Abstract**—The demand for data protection measures against unauthorized changes or deletions is steadily increasing. These measures are essential for maintaining the integrity and accessibility of data, effectively guarding against threats like ransomware attacks that focus on encrypting large volumes of stored data, as well as insider threats that involve tampering with or erasing system and access logs. Such protection measures have become crucial in today's landscape, and hardware-based solutions like Write-Once Read-Many (WORM) storage devices, have been put forth as viable options, which however impose hardware-level investments, and the impossibility to reuse the blocks of the storage devices after they have been written.
In this article we propose VaultFS, a Linux-suited file system oriented to the maintenance of cold-data, namely data that are written using a common file system interface, are kept accessible, but are not modifiable, even by threads running with (effective)root-id. Essentially, these files are supported via the write-once semantic, and cannot be subject to the rewriting (or deletion) of their content up to the end of their (potentially infinite) protection life time. Hence they cannot be subject to ransomware attacks even under privilege escalation. This takes place with no need for any underlying WORM device—since ValutFS is a pure software solution working with common read/write devices (e.g., hard disks and SSD). Also, VaultFS offers the possibility to protect the storage against Denial-of-Service (DOS) attacks, possibly caused by un-trusted applications that simply write on the file system to make its device blocks busy with non-removable content.
We have evaluated VaultFS in real-world scenarios and believe that it is an effective defense against ransomware, and more generally against data corruption attacks. representing an ideal solution in terms of security of cold-data via the write-once semantic, while keeping deployment and maintenance costs low.

## 1 INTRODUCTION

MANY critical applications require unalterable recording of data while ensuring continued accessibility. This need is particularly pronounced in sectors such as surveillance (where video and image streams are continuously archived), financial institutions, lottery results, court transcripts, regulatory filings, and so on. These records hold paramount importance in the context of compliance and audit requirements, with any unauthorized changes or deletions potentially carrying severe consequences.

Simultaneously ensuring data integrity and availability/accessibility presents a complex challenge, especially in face of powerful adversaries. In particular, the evolving landscape of enterprise ransomware threats has introduced a new layer of complexity. Modern ransomware groups have advanced their tactics, now targeting high-profile entities such as corporations and institutions, where the potential for substantial ransoms and damages is much higher [1] [2]. This trend renders conventional cryptographic data integrity methods, like digital signatures, less effective, as ransomware threats often involve the permanent encryption of valuable data, with access restoration being possible possible only via a decryption key hopefully provided upon a ransom payment.

To combat the spread of ransomware, the literature has proposed several techniques, which have been based on various approaches and methods. A few proposals rely on detecting if a ransomware attack is taking place and on attempting to block as soon as possible any action that can lead to additional/permanent damages in the file system (see, e.g., [3]). These solutions cope with "hot data", which are accessible for both read and write (update) operations by the active applications. Other proposals are instead based on the exploitation of the file system configuration and of its features in order to provide security support for stored files that should not be further modified. This takes place either via the setup and exploitation of file attributes [4], like data immutability, or via the reliance on specific instances of file systems which are made accessible according to strict protection rules and in limited operating conditions [5]. These solutions are suited for managing "cold data"—like critical backups—which once written should no longer be modified. They can be exploited for resuming the correct operations of the applications, even after a ransomware attack and according to a Recovery Point Objective (RPO) determined by the criticality of the applications.

In this article we focus on "cold data" protection, and provide a solution against ransomware attacks which is different from, and orthogonal to, what already proposed in the literature. In particular, we present an implementation of a file system for Linux, that we refer to as VaultFS, which

- *Pasquale Caporaso is with the Dipartimento di ingegneria Civile e Ingegneria Informatica (DICII) and CNIT Natl. Network Assessment and Monitoring Lab, Università di Roma Tor Vergata. E-mail: pasquale.caporaso@cnit.it*
- *Giuseppe Bianchi is with the Università di Roma Tor Vergata. Email: giuseppe.bianchi@uniroma2.it.*
- *Francesco Quaglia is with the Dipartimento di ingegneria Civile e Ingegneria Informatica (DICII), Università di Roma Tor Vergata. Email: francesco.quaglia@uniroma2.it.*

operates as a write-once register. In our file system, a file can be updated according to a sequential style (no seek) along a single I/O session. Hence, its content can never result as the encryption of some content already present into the same file. At the same time, once a file exists, no additional write session can be activated on it, making the file content usable for read-only accesses. Contextually, no modification or removal of the file data can occur via any operation by whichever (effective root-id) thread at the level of the block device keeping the file system. In particular, any attempt to operate at the block level on the device (or partition) that hosts VaultFS is intercepted and blocked along any time frame during which the file system is mounted. At the same time, we configured a facility that enables the file system to be unmounted only at shutdown of the operating system kernel. This still avoids the possibility for an attacker to damage the file system through operations at the device level via privilege escalation after an unmount of the file system.

VaultFS offers the write-once support at a pure software level, while still enabling the usage of read/write devices (e.g., hard disks and SSD). Hence it does not require Write-Once Read-Many (WORM) devices for supporting data immutability, like the ones targeted by the Microsoft's Project Silica [6], or those required by WORM specific file system implementations [7]. This makes our solution more widely usable considering common and low cost off-the-shelf device technologies.

Also, it is fundamental to note that saving critical data—like database backups and virtual machine images—on VaultFS provides a security level completely different from what could be offered by a classical file system, even under strict usage and configuration [5]. In fact, keeping this file system still accessible online (e.g. beyond a VPN) makes it again potentially tampered via file encryption by an attacker via privilege escalation. Contrarily, with VaultFS the write-once file-management semantic is guaranteed to be supported independently of the capabilities of running/-subverted applications. The way of operating of VaultFS is therefore substantially different from techniques based on making a file non-modifiable, for example by setting the "+i" flag/attribute via the `chattr` command. In fact, with these solutions, the file is still subject to modifications, and to data encryption, by a ransomware attack that is able to remove the flag via privilege escalation.

Its write-once nature and the possibility to keep files still accessible (while being no longer modifiable, even under privilege escalation), make VaultFS immediately adoptable for managing actual data (not just backups) in applications that produce them once, and then re-access data for read operations only, according to a classical read after write approach. Among them we can mention healthcare applications, whose reports (like MRI ones) have recently become principal objectives of ransomware [8]. But anyhow we can also consider common log files, whose content can have a critical role in scenarios where anomalies in the operations of applications (like security incidents) are identified, and need to be investigated relying on log data that are guaranteed to be correctly kept in the file system. In relation to these scenarios, VaultFS changes the concept of cold-data management in modern systems and applications, since it ensures write-once data survival in the presence of ransomware with no need for any additional data backup and without requiring any specific hardware-level support (e.g. WORM devices).

Additionally, file removal/rewrite in VaultFS is configurable, and can be setup to occur after a predetermined amount of time, making the storage reusable as well. We recall that this feature is not supportable when ensuring data integrity using classical WORM devices (e.g., CD-Rs, DVD-Rs and Blu-ray discs). Overall, each file that has been produced in VaultFS via the write-once mechanism remains permanent for a time period that can be configured when the file system is mounted. Once selected, this parameter value is not modifiable, even under scenarios of privilege escalation by an attacker.

Because of its core features, VaultFS could be ideally subject to Denial-of-Service (DOS) attacks leading to write-once operations of useless files, that in their turn keep the device storage busy with no possibility to remove them before the selected protection lifetime ends. To cope with this problem, VaultFS also offers an additional mechanism that enables the runtime check of the programs or code blocks that try to perform a write operation on a file. This enables protecting against DOS and further makes VaultFS suited for usage in scenarios where the applications that can actually write data on it (e.g. database backup tools like `mysqldump`) are predetermined.

VaultFS does not require any recompilation of the Linux kernel. Rather, it is all embedded within a Linux Kernel Module (LKM), hence fully relying on the last generation support that Linux already offers for inclusion of new operations at the kernel level.

In this article, we also report data related to the performance achievable via VaultFS when compared to Ext4 and Ntfs file systems, and we show a very negligible impact. Additionally, we demonstrate the compatibility of VaultFS with many applications oriented to the production of cold data, like common backup tools, as well as video surveillance applications, selected as a use case in the experimental assessment.

The remainder of this article is organized as follows. In Section 2, we discuss related work. VaultFS is presented in Section 3. Its experimental assessment is illustrated in Section 4. Conclusions are discussed in Section 5.

## 2 Related Work

Ransomware has been dealt with in the literature according to different approaches, also depending on whether protection involves hot (read/write) or cold (read after write) data.

As for hot data, the literature offers methods targeting the identification of the presence of malware, in order to attempt an early stop of its activity [9], [10], [11], [12], [13], [14], [15], [16]. Even though proposals specifically suited for the real-time identification of ransomware have been proposed [17], one of the main limits of these works is the response time, which can still lead to partial encryption of contents. Also, while most of the above cited solutions have performed experimental assessments demonstrating their ability to identify existing ransomware samples, the possibility of false negatives on the medium/long term (e.g.

in face of changes and evolution of ransomware attacks) is a standing problem.

Some authors have examined the use of machine learning as a defence based on the identification of the presence of attacker software [18], [19], [20], [21], [22], [23]. These works exploit different features in order to identify the attack, like for example API calls or sequences of opcodes in the software structure. In any case, these types of solutions need time from the collection of information in order to detect the attack. This might still lead to partially encrypted data on the file system.

An orthogonal approach for fighting ransomware is the one of relying on recovery techniques, which should provide the ability to recompute the original file content even after an encryption by a malware [10], [24]. However, even looking at the more advanced solutions, this way of proceeding is still linked to the possibility of false negatives—in fact, data for the recovery support are maintained until threads are still under judgement of their activities, and in any case not beyond a time limit.

Still for data recovery, the solution in [25] is an alternative proposal that the literature offers against ransomware. It is based on the recording of cryptographic keys managed at the level of the libraries hosted by the operating system, and on the exploitation of these keys whenever files need to be recovered after an encryption by ransomware. The main limit of this approach is that malware can include cryptographic services that allow bypassing operating-system hosted libraries.

The very recent approach in [26] presents a solution where file system updates are left pending into volatile memory, at the level of the page-cache of the operating system. They are finally flushed to the device (making the original file content overwritten) only after a delay used to more effectively assess if the threads that have written the new file content can be considered non-malicious. This is done relying on the acquisition of statistics related to the thread behavior along a time window. However, this solution can be still affected by false negatives.

Compared to all the above works, our proposal, which is suited for cold data, offers an orthogonal way of protection. Therefore, it could be combined with any of the aforementioned techniques.

As for literature solutions specifically oriented to cold data, we find [5]. In this work, the authors exploit a combination of techniques in order to provide a safe backup service. The cold copies of data (the backups) are kept in a separate partition of storage, and are made accessible exclusively through a virtual machine that is configured to host and run a minimal set of applications and services (those needed for the management of the backups). The main limit of this solution is that backups are anyhow maintained into a file system offering conventional data access support. Therefore the solution is bypassable in an attack scenario based on privilege escalation leading to root-level operations on the file system (and on the partition) devoted to the backups. VaultFS fully avoids this problem, since it ensures the write-once semantic independently of the capability level exploited by the running (malicious) software.

Various solutions offering the support for preventing deletion or change of data, and for meeting regulatory compliance requirements, have been based on WORM device technologies [6], [7], [27]. As noted, VaultFS is fully orthogonal to these solutions since it works with common read/write devices (e.g. hard disks and SSD), thus requiring no investments at all on specific hardware technologies. Also, being it a full software solution, it can operate in both bare-metal and virtual environments, still with no need for hardware specific facilities within the underlying platform. Furthermore, as we mentioned, VaultFS can be configured to manage a protection lifetime, after which the storage blocks originally used for a file are reusable, which is instead not directly allowed with common WORM devices. Hence, it provides a higher flexibility also in terms of actual usage of the storage. Additionally, it offers the support for DOS avoidance, in terms of usage of the device storage by non-trusted applications that simply try to saturate the file system with dummy data. This facility is not the target of WORM compliant storage systems.

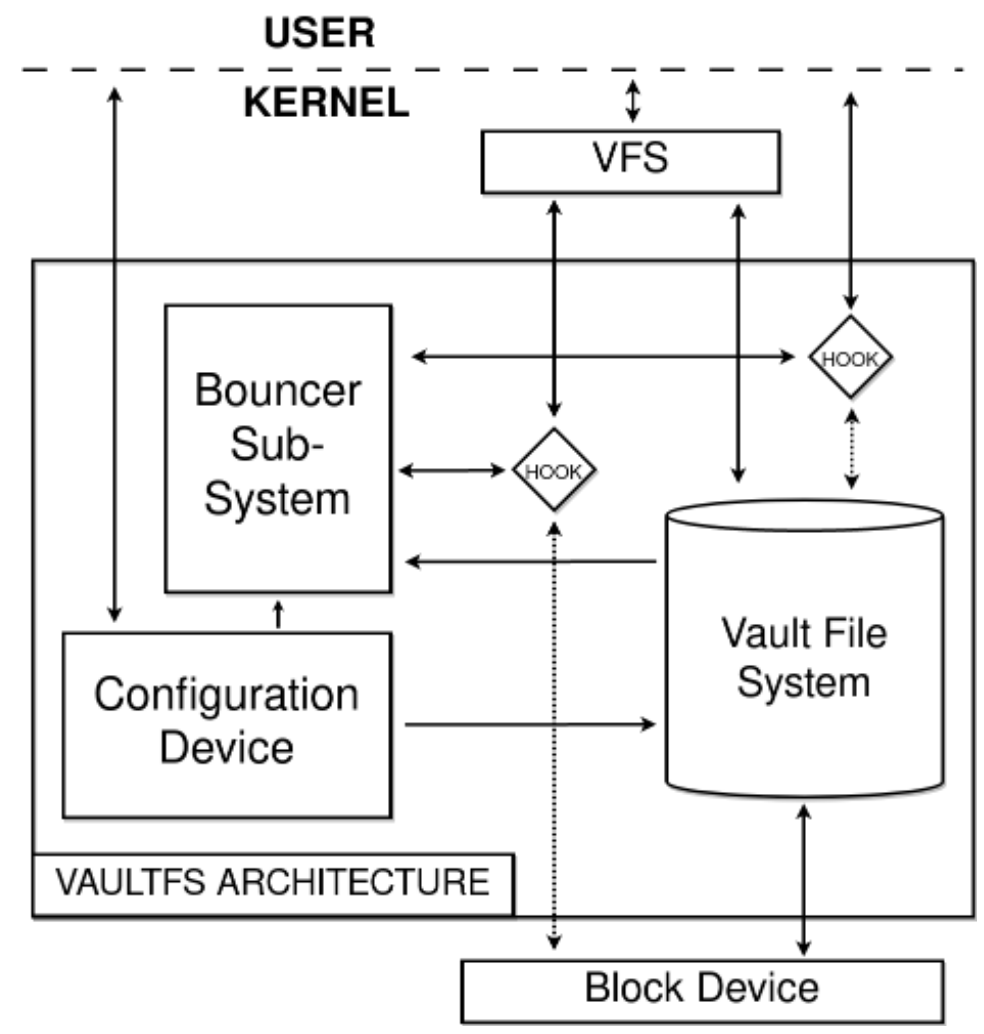

Fig. 1: Architecture of VaultFS.

## 3 VAULTFS

### 3.1 Baseline Concepts and Architecture

Working at the file system level in Linux requires considering the strict relation that various kernel-level subsystems have with the file system driver. In particular, it is typical that the parts of the driver of a specific file system (like the file operations it offers) exploit—or are exploited by—kernel-level services which are components of the more ample infrastructure of the Virtual File System (VFS).

The interactions with services that are "external" to the file system driver leads to the impossibility of guaranteeing the security levels we target by purely working inside the driver. Hence, our objective has been the one of constructing a comprehensive architecture surrounding the file system driver for guaranteeing that the security policies in place cannot be circumvented by malicious actors.

At a high level, the overall architecture of VaultFS can be schematized as shown in Figure 1. It includes three components, all of which can be added to the Linux kernel by relying on common APIs that Linux makes available for injecting any Loadable Kernel Module (LKM). Hence, all all our design is purely based on the LKM technology. The overview of these components is as follows:

- The *Vault File System* component is our file system driver. It takes care of managing files according to specific policies that are applied to any I/O session which can be ever opened by any process. In particular, it implements an Ext4-style file system, which only allows file writes along a single I/O session without overwriting—hence ensuring the write-once semantic. This feature fundamentally eliminates the possibility of data tampering while utilizing the file system, which is supported also via the embedding of admission control of operations working at level of the file system driver. Nevertheless, despite these limitations, this file system driver is compatible with most backup software and is fully adequate for other scenarios involving cold data management. As a last note, the *Vault File System* implements admission control mostly inside the different file-system specific modules it provides, and uses the kprobe Linux support for nesting hooks just to manage memory-mapping operations involving the page-cache kernel subsystem.
- The *Bouncer Subsystem* provides its support for the admission control of operations supported externally to the *Vault File System* driver, still by relying on hook functions nested via the kprobe support. In particular, it prevents the unmounting of VaultFS, and denies write-access to the underlying block device.
  As shown in the picture, the *Bouncer Subsystem* also offers hooks for kernel level functions that are external to the VFS. In particular, it includes hooks for the management of the memory-mapping kernel-level function, which is used to setup the structure of the address space of the active processes. These hooks enable the *Bouncer Subsystem* to check the content of executable pages into an address space, in order to determine if an active process that tries to write on VaultFS can be considered as legitimate. In fact, the *Bouncer Subsystem* also replies to queries that the *Vault File System* driver can issue. Through these queries, *Vault File System* can decide to accept or reject a write operation on a file, which is fundamental for enabling the file system to protect itself against the DOS attack.
- The *Configuration Device* enables configuring the VaultFS instance both at mount time and, if enabled by an administrator, also at runtime. In the latter case, it can be exploited through VFS system calls in order to enable changes on the admission of operations performed by both the *Bouncer Subsystem* and the *Vault File System* components. However, it is important to remark that this device is not a security-critical part of our architecture, just because it can be configured to deny at runtime any configuration modification, even if the `ioctl()` calls for interacting with the device are issued by (effective-)root-id threads. This is the default restrictive setup, which enables to achieve the maximum security level, despite privilege escalation. This default can be modified at the LKM load time just to allow for greater flexibility, and for enabling the end-user to exploit VaultFS according to his own needs.

As an important observation, the LKM that implements VaultFS does not offer unload functionalities. Hence, all the security features that are nested in the above three components, cannot be eliminated once the LKM has been loaded. At the same time, our solution deals with the implementation of a security oriented file system under the assumption that the Linux kernel is safe, and that attacks can only occur because of bugs or misscounfigurations at the user level. Hence, the kernel probes we install cannot be tampered by operations performed at kernel level.

Even though the Linux kernel is robust against attempts to perform non-trusted activities (like for example the discovery of the address of the struct kprobe table that embeds the information associated with a kernel probe, which can be exploited for tampering the probe), we know that having the possibility to mount any LKM allows performing any action that can tamper the kernel. This may for example happen under attacks based on privilege escalation. For extreme security-critical scenarios, the LKM that implements our VaultFS also offers the possibility to disable the init_module system call, by simply intercepting its activation, still via the kprobe service, and redirecting it to a non-regular execution that simply returns an error to the calling thread. Hence, we can disable the (malicious) mount of any additional LKM after the VaultFS one has been mounted. We feel this can be extremely useful, with no actual limitation, especially when exploiting VaultFS on a pure file-server system instance.

In the subsequent sections we present the details of each of the components being part of the VaultFS architecture.

### 3.2 The Vault File System

The centerpiece of our architecture is the *Vault File System*. It is a file system driver that we have developed in its entirety and is based on the layout and philosophy of Ext4. This file system driver is designed to host cold data and, as we pointed out before, it is founded on the ”write-once” principle.

There are some fundamental properties that must be satisfied via operations on this file system, which we list below:

1) A free i-node which can be used for a regular file creation is set as busy-protected via an `open()` system call that sets up an I/O session with write capability and that receives in input a file name representing a fully new hard link towards the i-node.
2) No additional I/O session with write capability can be opened on a busy-protected i-node related to a regular file, up to the end of the file-protection lifetime (see Section 3.4 for configurable exceptions). Hard links can still be set towards the busy-protected i-node.
3) Hard links towards a busy-protected i-node cannot be removed up to the end of the file-protection lifetime. Hence, the busy-protected i-node cannot become free up to the end of the file-protection lifetime.

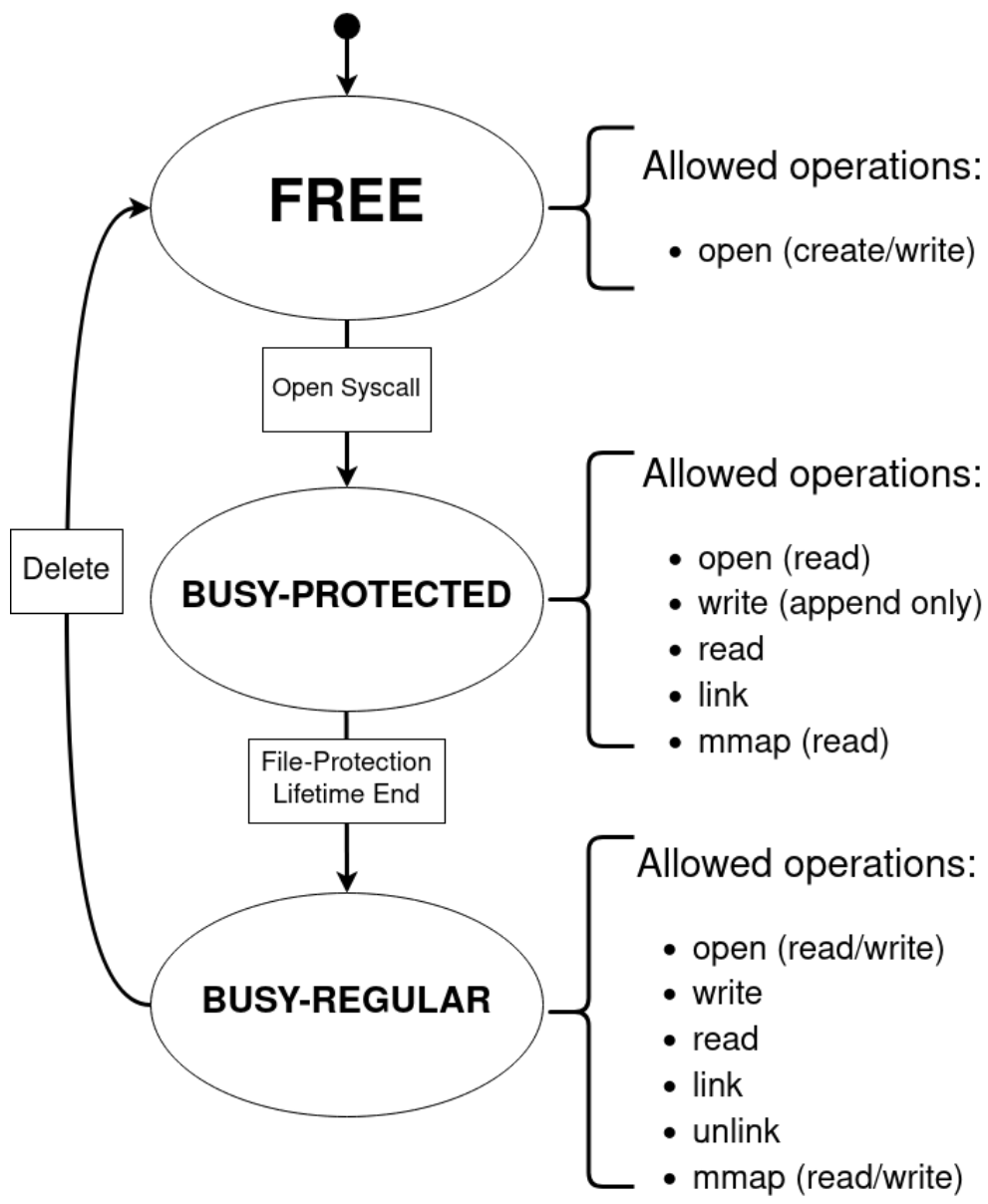

Fig. 2: i-node state machine

4) Once a write on a file whose i-node is busy-protected has been executed the written data must never be altered for the whole file-protection lifetime. This holds also for the unique I/O session with write capability we can have on a busy-protected i-node.
5) The above properties imply that data-blocks indexed by a busy-protected i-node cannot be released (and the file data they keep cannot be over written) for the whole file-protection lifetime.

For supporting all the above mentioned operations, we exploited a state machine, schematized in Figure 2, whose current state is kept in a flag into the i-node. At file system formatting time, each i-node not used for the root directory is set as free. When an `open()` system call with write capability is executed using a file-name not corresponding to any existing hard link, the file system driver finds a free i-node and puts it into the busy-protected state, updating the i-node flag. This prevents the file system driver from opening any further session with write capability for the whole file-protection lifetime.

Additional I/O sessions that only operate in read mode can be opened, either within or after the protection lifetime, and reading after a seek on any point of the file is supported in these I/O sessions.

As for files associated with directories (including the root directory), the *Vault File System* driver does not allow the removal of any hard link they keep up to the end of the corresponding file-protection lifetime (see point 3 in the above list). Hence any hard link to an i-node can be removed only after the i-node is no longer in the busy-protected state, i.e. it has been passed to the busy-regular state.

In order to control write operations when the i-node is in the busy-protected state (see points 4 and 5 in the above list), we leveraged the inherent behavior of conventional file systems. Upon the successful completion of a write operation, any file system driver will appropriately modify the size field within the i-node to account for the newly written data. We deem the write operation as committed at that particular moment and, consequently, regard all the data contained within the file up to that point as committed. This implies that any subsequent write can only be accepted if it does not modify any data existing from the beginning to the current size of the file. This check can be easily implemented at the start of the ”write” file operation inside the file-system driver. Furthermore, to adhere to the POSIX standard, we have modified the file operation ”lseek” to prevent for any i-node that is busy-protected the movement of the file offset to an area where write operations are not allowed.

However, we also need to consider that the file content in Linux is actually manipulated via the page-cache subsystem, which allows mapping page-cache pages into the address space of the applications. In particular this can occur when the `mmap()` system call is used for mapping a non-anonymous memory content into the address space; in this case one parameter that is needed for the system call is the file descriptor related to an open session on a file. This type of mapping is an additional channel an attacker might exploit to subvert the write-once semantic for files whose i-node is in the busy-protected state. In particular, even though there can be a single I/O session with write capability on a busy-protected i-node (see point 2 in the above list), the corresponding file descriptor can be exploited in an attack (also according to privilege escalation) in order to memory-map the file and rewrite its content. In fact, such a mapping would lead the application software to gain direct control on the file content, with possibility of performing updates even in the scenario of a file that has been just opened (created) and populated via a unique session with write capability.

To avoid this scenario, our solution exploits the kprobe facility offered by Linux to install a hook on the `mmap()` system call. The hook checks if the file descriptor passed in input corresponds to an i-node in the busy-protected state. In the positive case, the mapping service returns with an error code if the corresponding I/O session has write capability, preventing the mapping operation. At the same time, the operation is still supported with no restriction for files whose i-node has passed to the busy-regular state.

Still in relation to the management of i-nodes, we decided not to support the creation of char/block devices via `mknod()` in VaultFS, except if a specific parameter is selected at mount time of our LKM (see Section 3.4). The reason for this choice is related to the fact that any i-node for a char/block device can be associated to any char/block-device driver offered by Linux. This driver is external to our *Vault File System* driver, hence our software cannot provide any support for ensuring that specific properties, e.g. in terms of security and data durability, are actually ensured for a busy-protected i-node created for such char/block device.

A crucial aspect of the architecture is related to the notion of ”life time” of the i-node (or file) protection. This is the period of time an i-node needs to remain in the busy-protected state, and cannot be reused for keeping data related to any other file or for performing updates of data that reside in the file. Essentially, this life time corresponds to the duration of the insurance of the write-once property for files in the file system.

This life time can be ideally set as unlimited, which means that a file hosted by a mounted instance of the file system can be subject to no removal and no rewrite at any point in the future. This is the scenario where the state machine depicted in Figure 2 boils down to a two-state configuration since the busy-regular state of the i-node can never be reached along time. However, if we consider the case of exploitation of our file system for keeping backup data (e.g. database or VM backups to be kept accessible, although protected, for dealing with any critical scenario), the impossibility of adopting a finite life time of the protection could make the cleanup of no longer needed backups and data challenging.

To cope with this aspect, we exploited the notion of Time To Live (TTL), which can be configured at the file system mount and represents a time interval, starting from the file creation, after which the file is considered obsolete for what concerns write-once insurance, and, consequently, all (over)write and delete limitations (including the delete of hard-links) are lifted. After this TTL, the file i-node passes to the busy-regular state, and is managed according to the conventional (regular) operations any common file system, like Ext4, supports. This configuration offers a more straightforward maintenance process and, in general, increases flexibility, as the TTL can still be set to an infinite value for specific cold-data management scenarios—like the one where the file system is used for files representing MRI data in healthcare systems.

The implementation of the TTL feature requires in its turn some careful considerations in order not to become a critical point for security. In particular, in order to determine whether to lift the restrictions or maintain an i-node in the busy-protected state, it is crucial to possess a reliable method for measuring the passage of time within the system. However, the current Linux kernel APIs do not provide a suitable solution for this task. The real-time ktime accessors return a value that can be modified by both users and NTP servers [28], while the values provided by the boot time and monotonic clock are reset during each system shutdown. It is possible to intercept the shutdown process and update an internal counter to track the actual elapsed time, but this method would prove ineffective in cases of forced power-off such as power outages.

To completely resolve this issue, we have included in our module a kernel daemon in the form of a workqueue item, which is triggered at regular intervals. The only task of this daemon is to update an internal time counter known as the "file system time" (FST), which resides within the superblock of the *Vault File System* instance. The FST represents the actual time that has elapsed since that specific file system instance was mounted for the first time. We note how FST is active only along periods where the VaultFS in mounted, since its superblock is reachable via the function associated with our work-queue item only in this scenario. However, this is perfectly aligned with the objective of VaultFS, which has been thought as a storage system to be constantly kept mounted and active (for enabling the access to the recorded files), while still being protected against attacks like ransomware, even under privilege escalation.

In the current setup FST can operate at the granularity of an individual second, which is the update interval of the daemon, even though we suggest setting it to work with granularity of the order of 10 seconds, so as to actually achieve no interference at all with the operations of other workqueue-items the Linux kernel can setup. Also, given that this time measurement is specifically utilized for checking the file-protection life time, and for passing an i-node from the busy-protected to the busy-regular state, which typically spans periods like weeks months or even years, according to classical data-maintenance, higher precision (less than 1 second granularity of the timer) is deemed unnecessary. With this architecture, our file system ensures resilience to power-off events and system crashes as the value of FST is periodically saved to the disk partition that hosts the VaultFS. Even in the scenario where a shutdown occurs just before the daemon's time update, the discrepancy in the FST value would be less than the setup period (e.g. 10 seconds). Considering the context, we believe this error is negligible, furthermore, even in such a scenario, the FST would be lower than its theoretical correct value, resulting in a prolonged life time of file protection which means that there is no way this approach can be used for an attack.

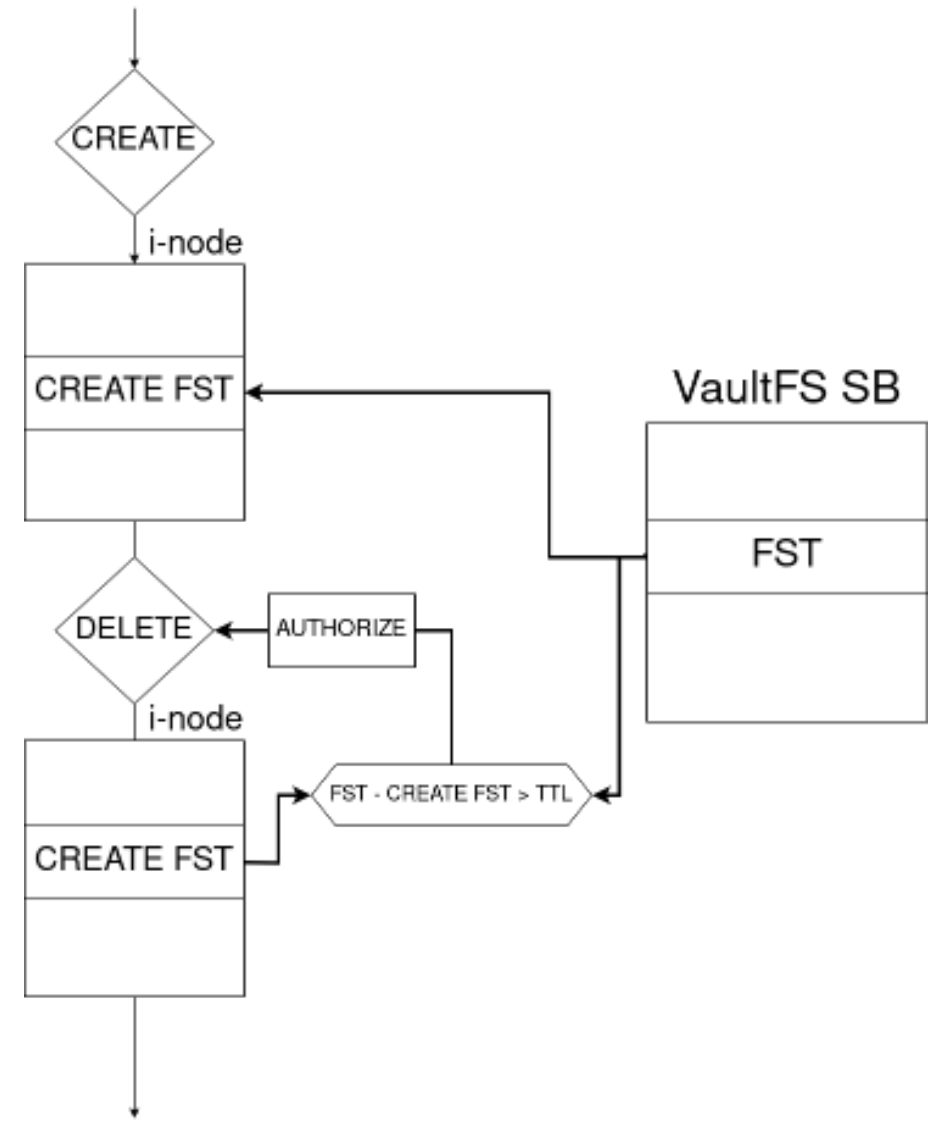

Fig. 3: Create and delete operations with FST

In Figure 3, we illustrate the usage of the file system time (FST) in two specific scenarios. Firstly, during the file creation process, VaultFS reads the current value of the internal time counter (FST) and stores it within the corresponding i-node of the file. Secondly, when a delete request is received, VaultFS retrieves the current value of FST, subtracts the saved internal creation time obtained from the i-node, and compares it against the TTL value set by the system administrator. If the calculated time difference exceeds the TTL, the i-node is passed to the busy-regular state and the delete operation is permitted; otherwise, it is denied.

In Section 4, we will demonstrate that the VaultFS architecture does not conflict with existing software tools and will not impede both the backup process, via classical backup programs, and the storage of data coming from video surveillance applications, which are exploited in this article as a use case in relation to cold-data management.

In any case, as we will show in Section 3.4, VaultFS also offers (1) a flexible mount scheme where options can be configured and (2) the optional support for runtime reconfiguration of its operations, which can be exploited in scenarios where extremely high flexibility is requested and specific trade-offs between security and the capability of supporting tasks is the objective for its usage.

### 3.3 The Bouncer Subsystem

The *Bouncer Subsystem* we included in our architecture has two main goals: firstly, to ensure that the only possible way to access data on the file system is through the *Vault File System* driver, and secondly, to ensure the up-time of VaultFS.

The first objective is self-evident: VaultFS requires exclusive access to the storage device to protect it with its restrictive write-once permissions, at least for files whose i-node is currently busy-protected. However, this is not typically the case in most modern VFS architectures, as access to the storage device is permitted not only from the file system driver, but also directly from the block device beneath it. To mitigate this vulnerability, the *Bouncer Subsystem* intercepts all mount events and records the major/minor numbers of all block devices hosting a VaultFS instance. Subsequently, as shown in Listing 1, when a user attempts to access a path using the `open()` system call, the *Bouncer Subsystem* intercepts the call, still via kprobe and verifies whether it is directed towards one of the previously recorded major/minor numbers and if so, the operation is denied.

```
//check if we are accessing a device directly
bdev = name_to_dev_t(path);
if (bdev) {
 //check if device is protected
 list_for_each_entry(info, &vaultfs_instances, n) {
  if (info->bdev_id == bdev) {
   //it is protected - fail
   res->pass = 0;
   break;
  }
 }
}
```

Listing 1: Limiting block device access

The second objective is multi-faceted. Keeping VaultFS active means not only guaranteeing the write-once semantic, but also that:

A) The file system is kept mounted, in order to protect the access to the block device according to what explained before (see Listing 1).
B) The file system storage is not subject to DOS. In particular, ensuring that, according to some capacity plan, it still has space to store data, could be central in automatic backup or data store procedures.

To cope with point A), the *Bouncer Subsystem* places a kernel probe on the `umount()` system call, which is shown in Listing 2, denying the operation if the object being unmounted is a VaultFS instance.

```
//pass by deafult
res->pass = 1;

//check if path is protected
list_for_each_entry(info, &vaultfs_instances, n) {
    if (strcmp(info->mount_path, path) == 0) {
     //is is protected - is the lock active?
     if (info->umount_lock) {
      //it is active - fail
      res->pass = 0;
     } else {
      //it is not - clean up the list - pass
      list_del(&info->node);
     }
     break;
    }
}
```

Listing 2: Blocking umount

Maintaining sufficient space on the device (see point B) presents a more complex problem. Due to the pseudo-permanent nature of the files in VaultFS, an attacker could employ a DOS attack by filling up the hard drive with useless yet still non-erasable files, rendering the system unavailable. The same is true for hard-links, kept into the directory files. To address this issue, we have incorporated a whitelist of programs that can be added during the file system mount by a system administrator. Any program not included in the whitelist will be prohibited from writing to the VaultFS instance, thereby preventing a DOS attack. The discovery on the possibility to write data is done by the *Vault File System* driver via queries issued to the *Bouncer Subsystem*. These queries are issued when the `open()` and `write()` system calls are handled, as well as for the handling of the `link()` system call, which manages the creation of hard-links.

The creation of the whitelist and the authentication process can be schematized as shown in Figure 4, and work as follows:

1) During the hard drive formatting a list of approved programs and libraries can be included. This list is stored in the file system superblock.
2) While mounting VaultFS, our software reads the list, identifies the files associated with the approved programs, and generates an hash for each file. These hashes are then stored in the file system superblock along with their corresponding absolute paths.
3) While VaultFS is mounted, whenever a new process maps a memory area associated with a file, we again intercept the mapping. If the path of the allocated area matches one of the authorized paths, we compute the hash of the area and compare it to the hashes stored in the superblock. If there is a match, the program is considered authorized. If there is no match, the program is permanently banned from accessing VaultFS in write mode, and any future memory mapping made by that program are not monitored.

The central aspect of this architecture revolves around the computation of hashes. The efficiency of this calculation is crucial as it is performed synchronously during memory mapping. Simultaneously, we have to ensure the security of the hash to prevent malicious attackers from easily by-passing the check. To strike a balance between speed and security, we have implemented the following rules:

- Any process with an associated filename that is not present in the whitelist is automatically banned without hash computation. This measure can effectively prevent the majority of memory mapping operations from triggering hash calculation.
- At file system mount time, instead of computing the hash for the entire file (program) that is in the white list, we divide the file into areas (sets of pages belonging to several portions of the executable section of the files), and compute the hash of each individual area. The size and position of these portions are determined using a random number based on a random seed determined during the file system mount process, which is specific to each VaultFS instance. This means that the hashes stored in the superblock change every time. The total size of each hashed area varies between 2 and 10 memory pages, depending on the size of the original file. When in a process this same file (program) is executed, and the mapping of its areas takes place, we randomly select what area(s) to hash and check against the originally computed hash values. To ensure security, the areas selected for the hash of each executed instance of a same program are as well determined randomly. This approach guarantees that an attacker cannot determine which specific area is being used for the hash and consequently cannot construct a malicious authorized binary. At the same time, our system hashes only one or a few randomly selected areas at each program execution (rather than all the program areas), reducing the cost of this operation.

This solution is based on the determination of the hashes of sets of pages as computed at the mount time of VaultFS. Hence, we prevent the inclusion in the white list of any program that includes memory zones (sets of pages) that are marked as write-exe, since their content can be changed at runtime.

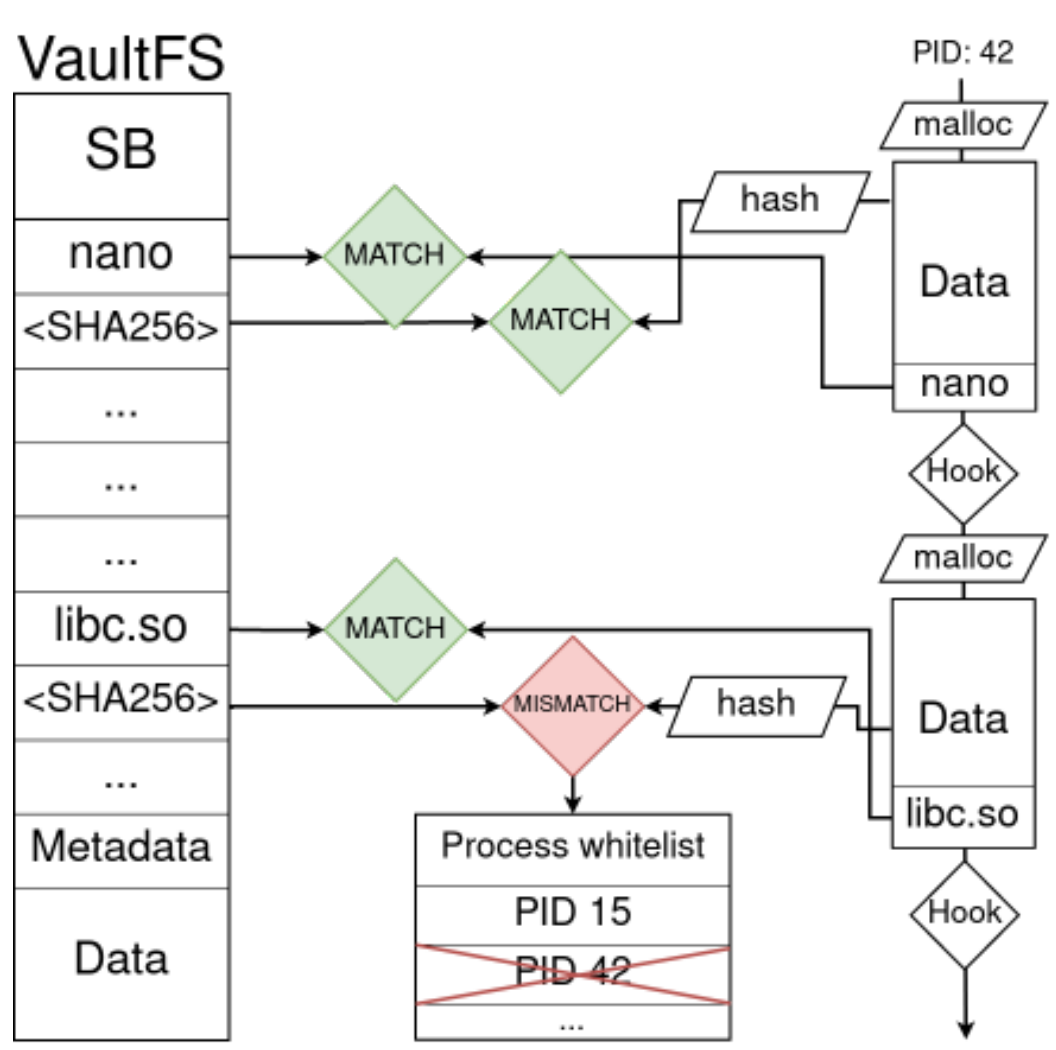

Fig. 4: Process whitelist example

| Option | Default setup |
|---|---|
| TTL | enabled |
| Append reopen | disabled |
| Program withelist activation | disabled |
| Char/block device creation | disabled |
| Communication device | disabled |

TABLE 1: Default setup of the configuration options of VaultFS

### 3.4 Configuration Device

To facilitate integration of VaultFS with existing infrastructures, we have incorporated a variety of configuration options that enable system administrators to customize the file system functionalities to their specific needs. Each option can be selected during mount by configuring the appropriate value in the file system superblock, with the configuration being specific to each instance, thereby allowing for different configurations on different partitions of the hard drives. The currently allowed settings include:

- *TTL*: This option defines the file-protection life time. If the option is not enabled at the file system mount, the value *"infinite"* is used as the default.
- *Append reopen*: This option determines if a file having its i-node in the busy-protected state, which is not currently opened in write mode, can be reopened in write mode for appending new data at its end. This parameter can be useful when VaultFS is exploited for securely keeping log files, which can be reopened in append mode by the applications. With VaultFS these files can still be kept on-line with no risk of tampering of their content via attacks, even under privilege escalation. This option is disabled by default.
- *Programs whitelist activation*: This setting activates the program whitelist as mentioned in Section 3.3. By default this option is disabled.
- *Mount unlock*: This option enables the system to be unmounted without shutting down the operating system, using a per-instance password. This option is disabled by default.
- *Char/block devices creation:* This option determines if the creation of char/block devices in directories of VaultFS is allowed. This option can be also configured in order to enable the creation after passing through a query to the *Bouncer Subsystem* in order to detect if the process that is calling the creation is whitelisted. By default this option is disabled.
- *Communication device*: This option determines if this instance of VaultFS needs to accept configuration directives at runtime, with the possibility of dynamically reconfiguring any of the above listed options. Also this option is disabled by default.

The default setup for all the listed options is summarized in Table 1.

For supporting the communication device option, the LKM of VaultFS includes a dedicated `ioctl()` device that can directly modify the superblock of the file system instances to change their behavior, after receiving directives from authorized users. An authorized user is identified with a per-VaultFS-instance password, chosen during mount, which cannot be changed and is securely stored inside the superblock. Using this device, a system administrator can change all the settings that were decided during mount. Also, we note that TTL changes are not retroactive, which means that existing files will maintain their configuration as it was when they were created, this implies that files with different configurations can coexist on the same VaultFS instance, and changing the configuration will not lower the security of already written files.

To implement this functionality, our system includes a global list within the LKM which maintains information regarding all mounted VaultFS instances. When a new instance is mounted, it registers its details in the list, making them accessible to other subsystems. Specifically, the information includes a pointer to the in-memory superblock structure, which can be accessed from the communication device to allow for any necessary configuration changes, if required.

### 3.5 Security Evaluation

Considering everything that has been explained, our system is capable of guaranteeing file integrity (i.e., no rewriting or deletion of its content) for the whole file-protection lifetime, when the setup ”Char/block device creation” and ”Communication device” is compliant to what specified in Table 1, under the following threat model:

A) The attacker is able to execute commands/software with root privileges on the machine.
B) The kernel and drivers of the machine are secure, i.e., there is no bug that could cause arbitrary code execution in the kernel.
C) Our file system is mounted during the boot process before any attacker is able to inject commands.

We believe that this threat model covers an overwhelming majority of security incidents. Point B is generally true on most stable Linux distributions, and as we discussed in Section 3.1, VaultFS offers the support for disabling the loading of additional LKM modules which could somehow (dynamically) lead to weaker kernel configurations. Point A represents the highest possible amount of power that one can give an attacker (against a secure kernel), and, unlike many other security solutions, our system remains effective under this threat.

Finally, there are various approaches that can realize point C. The responsibility of mounting our file system could be entrusted to an external network controller, which verifies the online status of our system before restoring external connections. Also, this point is automatically guaranteed in systems that use PXEs[1] [29], which are prevalent in modern data centers [30]. In such setups, the operating system image and its configuration are centrally maintained on a remote server and cannot be manipulated by an attacker to execute commands before file systems are mounted.

## 4 Experimental Assessment

In this section, we provide an experimental assessment of VaultFS. We consider different aspects, which allow testing both its usability and performance, in particular when comparing it against an instance of Ext4 or Ntfs.

All the tests we describe have been carried out on a machine equipped with an Intel i7 processor running at 3.1 GHz and 12 GB of RAM. The instances of VaultFS, Ext4 and Ntfs, which we compared with each other for what concerns performance, were all kept on identical 32 GB USB drives.

As we noted, the LKM module supporting VaultFS has been designed to ensure compatibility with the latest Long-Term Support (LTS) version of Linux distributions. In this study we relied on version 22.04 of Ubuntu, which incorporates kernel version 5.19. Also, we configured Ubuntu with an Ext4 root file system.

### 4.1 Compatibility

VaultFS has been specifically designed for storing and managing cold data. Considering this target, we concentrate on assessing the compatibility of VaultFS with established software in the backup and video surveillance domains. In any case we recall again the ample set of use cases that VaultFS has, like for example the maintenance of medical test results. Applications like the latter mentioned one have not been considered in this study simply because of the unavailability of proprietary software, the use of which would have been necessary for the experimental phase.

#### 4.1.1 Backup

In order to evaluate the compatibility of VaultFS with backup tools, we selected widely used software solutions for both file system and database backups. These tools have been used to perform a backup operation from an Ext4 file system to a VaultFS instance. We configured VaultFS according to both the default setup (denoted as ”Default Protection”) indicated in Table 1, and with the ”Append Reopen” option activated. In any case, the life time of the file protection has been set to one year. We recall that under both the used setups of VaultFS, there is no possibility to rewrite any single byte of a file (or a directory file) for the whole protection life time, even when threads run with (effective)root-id capabilities.

| | Append Reopen | Default Protection |
|---|---|---|
| GUI Copy Paste | ✓ | ✓ |
| rsync* | ✓ | ✓ |
| flexbackup | ✓ | ✓ |
| unison* | ✓ | ✓ |
| mysqldump | ✓ | ✓ |
| Deja-dup | ✓ | ✓ |
| Kbackup | ✓ | ✓ |

TABLE 2: Compatibility with backup tools.

The findings of this experiment are presented in Table 2. The results clearly indicate that VaultFS is compatible with all the tested backup solutions, across both the tested configurations, without requiring any modification to the default backup-tool options. However, it should be noted that for the UNISON and RSYNC tools, an additional flag is necessary to prevent folder renaming during the copy process, as this operation requires the elimination of a hard link, which is not permitted by VaultFS for the whole protection lifetime of any file (including directory files).

1. Preboot Execution Environment

#### 4.1.2 Video Surveillance

Similarly to the backup compatibility experiment, we conducted a test with video surveillance software. We selected multiple commonly used software solutions in this field, which differ in terms of how the video data is stored and the configuration settings associated with their setup. For this experiment, we configured each software to save the video or photo data captured by an USB webcam onto a VaultFS instance. At the same time, we made each software tool configuration files (or software specific files) reside on the root file system instance (as we mentioned this is an instance of Ext4) mounted by Linux, as it typically occurs for most of the applications. Also in this case we executed the tests by considering both the "Default Protection" offered by VaultFS and the activation of the "Append Reopen" option. In both cases we configured one year life time of the file protection.

| | Append Reopen | Default Protection |
|---|---|---|
| ZoneMinder | ✓ | ✓ |
| Webcamoid | ✓ | ✓ |
| ivideon | ✓ | ✗ |
| xeoma | ✓ | ✓ |

TABLE 3: Compatibility with video surveillance tools.

The results of these tests are summarized in Table 3. As we can see, almost all the software tools have been successfully used with VaultFS in both its configurations ("Default Protection" and "Append Reopen"). The unique tool that required the "Append Reopen" setup for being correctly used is IVIDEON.

### 4.2 Performance

Ensuring an acceptable level of performance is crucial for all security solutions. In this section, we report data for the assessment of the performance impact of VaultFS and of its different components. Our objective has been to show that no component of the architecture would introduce any substantial or unacceptable performance overhead.

#### 4.2.1 File system statistics

VaultFS, despite being based on the Ext4 file system, is still a fully new file system, because of its peculiarities in the management of i-nodes and file data. Therefore, as an initial test, we aimed to show that its performance is comparable to other commonly used file systems in the Linux world. To assess this, we performed a file copy from/to each of the file systems we are comparing, namely Ext4, Ntfs[2] and VaultFS. As mentioned, these are installed on identical 32GB USB hard drives. The objective of these tests has been measuring the time taken to complete the file copy operation for files of various sizes, ranging from megabytes to gigabytes. Each value we report is the average over 100 samples, and we also report the standard deviation over the gathered samples.

The results of this test are presented in Figure 5. They show that VaultFS outperforms Ntfs, and that it shows a limited performance drop compared to Ext4, which tends to appear when considering larger files. In particular, for 3GB file size, VaultFS is no more than 10% slower than Ext4. Considering that this result is obtained relying on the simple CP program, which essentially performs a continuous iteration of read/write operations from/to source/destination files, this test represents a kind of worst case scenario for the impact of VaultFS on threads' activities. In fact, the thread that is executing the CP program does not experience any significant latency caused by software components that are outside the VFS architecture of Linux, which would have reduced the relative cost/overhead of file system operations[3]. In any case, VaultFS can be subject to fine tuning and additional software optimization along time (like pre-reserving of blocks for new file data), specifically linked to performance improvements.

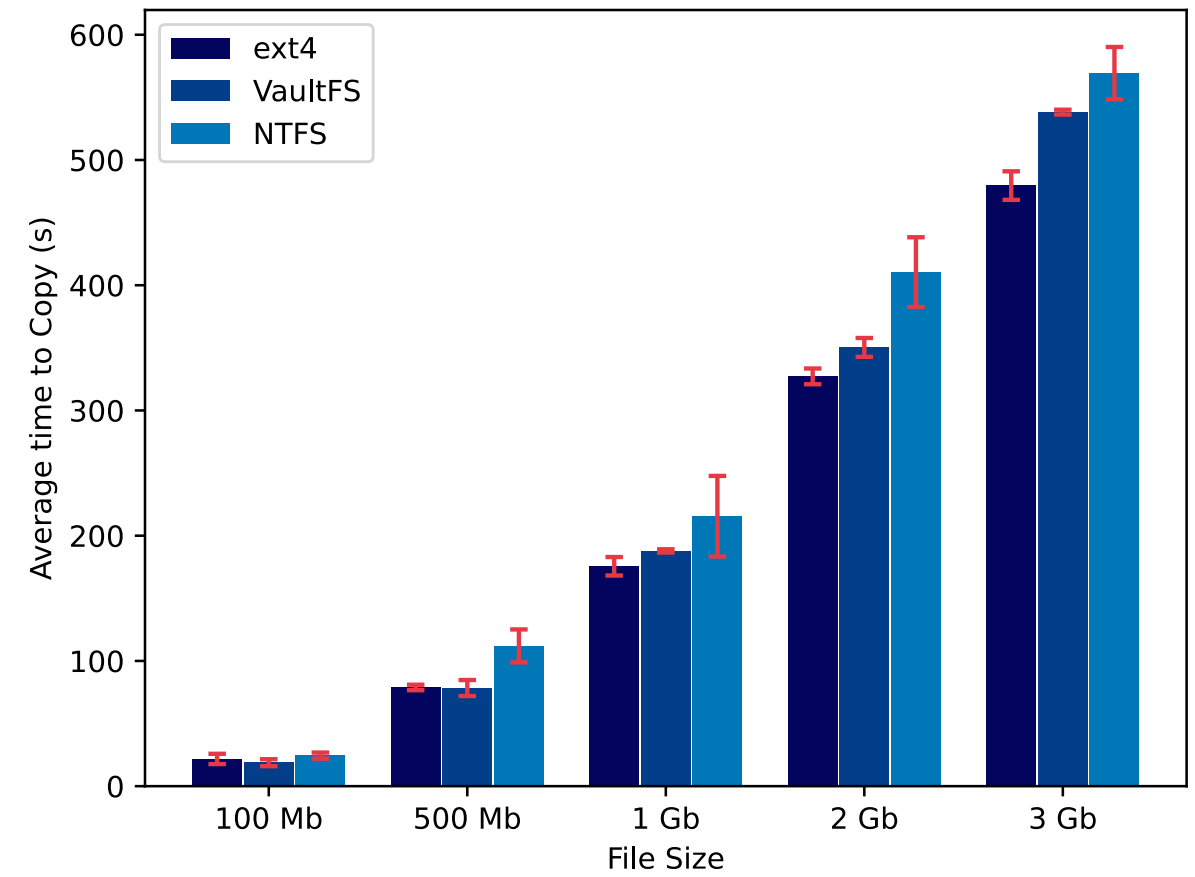

Fig. 5: Average copy time.

#### 4.2.2 Open kernel-probe slowdown

As we have explained, the *Bouncer Subsystem* requires a kernel probe on the `open()` system call, which is frequently invoked during normal system usage. Hence, it is crucial to ensure that our probe does not significantly impact the execution time of this system call. To assess this, we conducted multiple open operations on an Ext4 file system instance, still hosted on the aforementioned 32GB USB hard drive, measuring the time taken to complete them both with and without the VaultFS LKM loaded.

The results, categorized by path depth, are illustrated in Figure 6. The value zero for path dept indicates that the `open()` system call opens a file in the current Process Working Directory (PWD). Each reported value is still the average of 100 runs, and we also report the standard deviation. We would like to note that the values provided were determined under the assumption that the path being opened is absent from the page-cache. To ensure this condition, we unmounted the file system after each measurement. Notably, there is no discernible slowdown caused by our module, indicating that the execution time of the `open()` system call remains largely unaffected when it becomes

2. We used the Linux Kernel implementation of the Ntfs file system, contained in the */fs/ntfs* folder of the source code.

3. The thread running the SC command can experience a delay related to the materialization in RAM memory of the virtual pages it is using for hosting data read from the source file, but this is somehow negligible considering the iterative structure of the program.

blocking since there is at least one element of the path (e,g. the specific file name) whose i-node needs to be read from the USB hard drive.

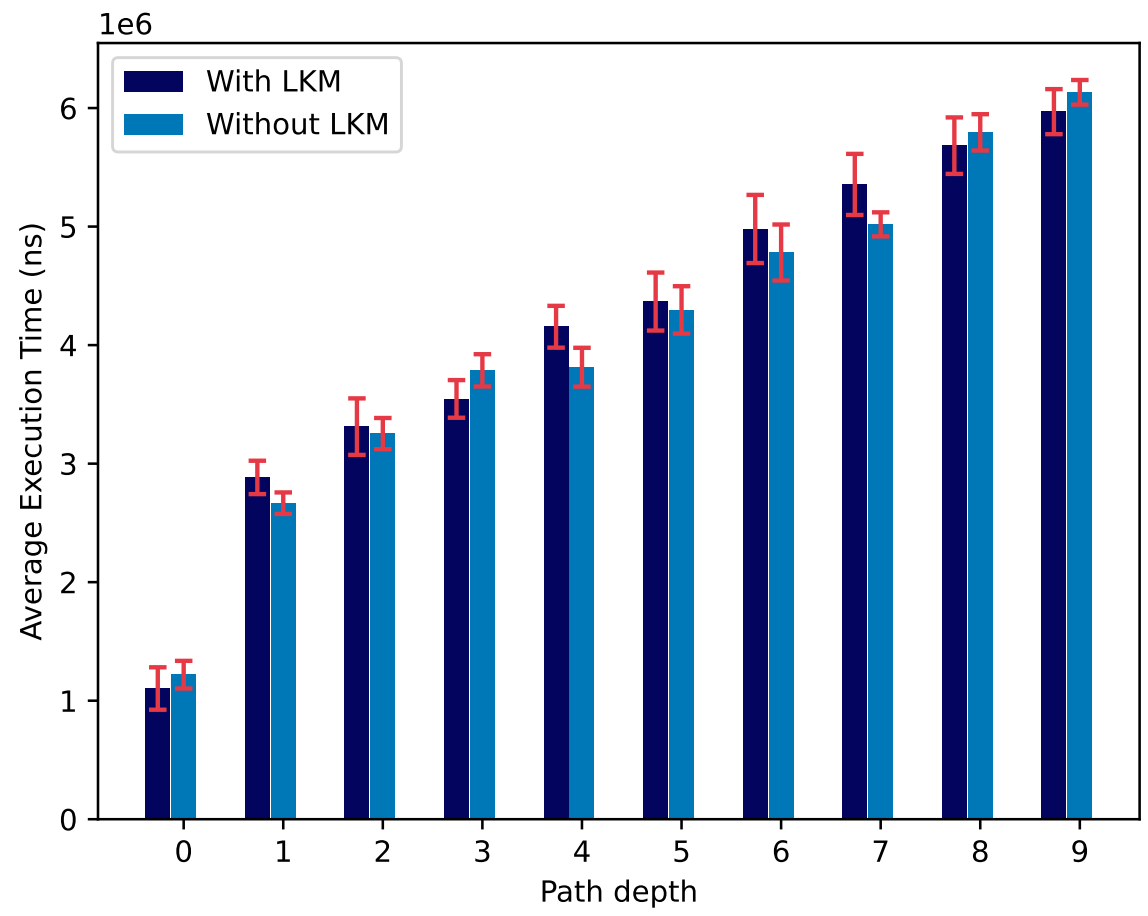

Fig. 6: Average time for opening a file.

In any case, for completeness we also report in Figure 7 data related to the opposite scenario where all the items in the path are already kept in the page-cache. This time the actual execution of the thread opening the target file is essentially non-blocking, and we have the maximum impact of the kernel probe we installed on the `open()` system call. In this case, the `open()` operation when the VaultFS LKM module is not loaded is approximately up to 40% faster compared to the scenario where the path elements were not cached. Nonetheless, considering the data in Figure 6, we have that a single path element that is not present in the page-cache leads to the same latency values for opening files when the LKM of VaultFS is loaded or not. This leads to the outcome that the scenario for which an opening delay is truly affected when the LKM of VaultFS is loaded is limited to the (somehow unlikely) case where no element in the path is out of the page-cache. This can have a reduced statistical incidence. Furthermore, considering the exceptionally rapid completion of the operation when everything is cached, we believe that the absolute slowdown introduced by our system is negligible.

#### *4.2.3 Process Whitelist slowdown*

The whitelist facility likely introduces the most significant performance overhead in VaultFS, as it adds operations during every memory mapping event while managing the address space of a process. To assess the acceptability of this performance overhead, we conducted several tests. The objective was to measure the execution time of multiple command line utilities in three scenarios:

- normal environment (without VaultFS);
- with VaultFS mounted and without the command line software in the whitelist; and
- with VaultFS mounted and with the command line software in the whitelist.

Furthermore, as explained in Section 3.3, we have configured VaultFS to hash different percentages of the process address space and measured how this affected performance. It is important to note that higher percentages of hashing represent a stronger support to security against DOS, but also introduce more performance overhead. The results are depicted in Figures 8 and 9, and the slowdown values are reported in Table 4.

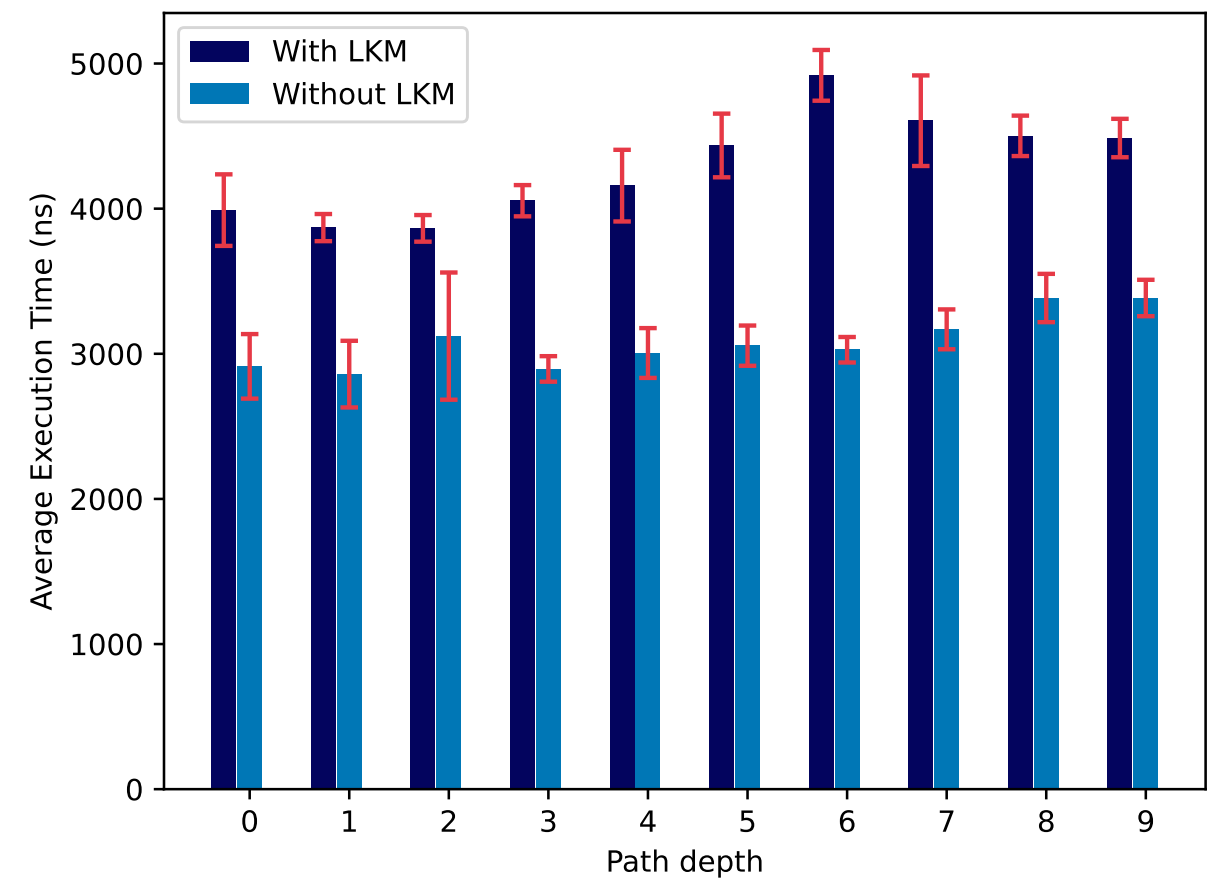

Fig. 7: Average time for opening a file with all path structures in the page-cache.

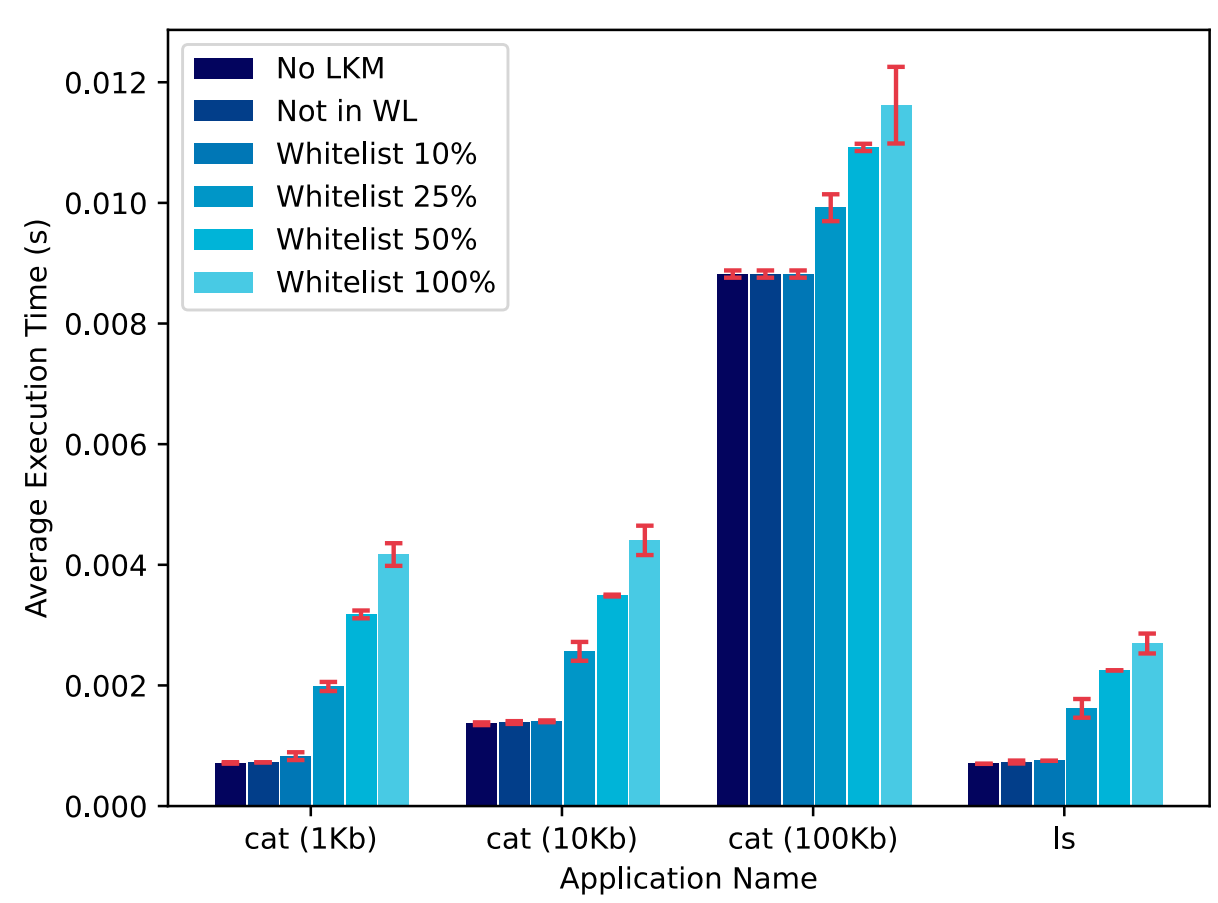

Fig. 8: Average execution time for short-lived command line software — less than 0.01 seconds per execution in the normal environment (average and standard deviation over 2000 executions).

We have categorized the values based on the average execution time, considering any program with an average execution time of less than 0.01 seconds in the normal environment as a short-lived application. We can observe the following trends:

- When the application is not in the whitelist, the slowdown is negligible ($< 1\%$).
- When hashing 10% of the address space, we observe a low performance overhead ($< 16\%$ slowdown) for all applications.

| | rsync (1Kb) | cat (1Kb) | cat (10Kb) | cat (100Kb) | zip (1Kb) | zip (10Kb) | ls |
|---|---|---|---|---|---|---|---|
| Not in Whitelist | 0% | 1,4% | 1,5% | 0% | 0% | 0,7% | 4% |
| 10% Hash | 0% | 15% | 2,9% | 0% | 0,4% | 4,7% | 7,1% |
| 25% Hash | 7,4% | 173% | 87,9% | 12,5% | 7,4% | 1,3% | 131% |
| 50% Hash | 12,7% | 344% | 155% | 23,8% | 8,9% | 1,4% | 221% |
| 100% Hash | 23,5% | 483% | 227% | 31,7% | 13,5% | 2% | 285% |

TABLE 4: Slowdown for common software.

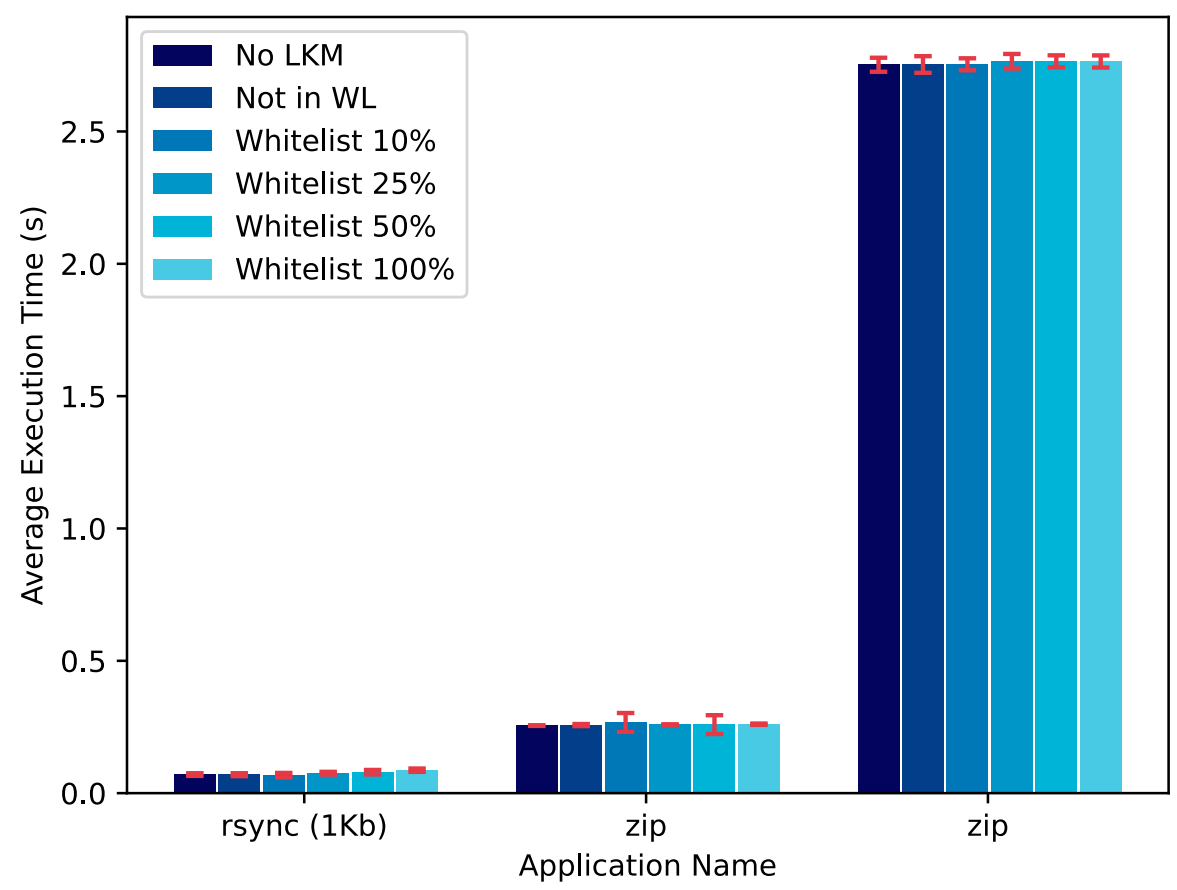

Fig. 9: Average execution time for long-lived command line software — more than 0.01 seconds per execution in the normal environment (average and standard deviation over 500 executions).

- When hashing 100% of the address space, long-lived applications do not experience any significant slowdown ($< 24\%$).

Given the very low delay representing the barrier between short and long lived applications, the data show how our solution is essentially non-intrusive under all the scenarios where a non-minimal amount of work is requested for managing/storing data. At the same time, for scenarios where the required amount of work is minimal, the overhead introduced still brings to very small turnarounds of the applications. Additionally, if the performance overhead for short-lived applications is a concern, VaultFS can be set at the lowest hashing percentage of the application address space, i.e. 10%, which still protects against DOS while resulting in a negligible slowdown. Additionally, all applications not included in the white list will experience no slowdown at all, which supports the pragmatical usability of VaultFS in contexts where both whitelisted and non-whitelisted applications can be simultaneously used.

## 5 Conclusions

Data survival on file systems has become one of the major security problems, especially when considering ransomware malware, which crypts data making them re-accessible only after a ransom is payed. In this article we tackled this problem and proposed VaultFS, which is an implementation of a Linux-tailored file system that supports the write-once semantic of files. The peculiarities of VaultFS are that 1) it guarantees file content immutability even against attackers that work with (effective)root-id, and 2) it ensures data immutability while working with common off-the-shelf read/write storage (e.g., hard disks and SSD), thanks to an articulated software architecture. Also, VaultFS has a design perfectly aligned with the facilities Linux offers for setting up kernel subsystems, since it is fully implemented by a Linux Kernel Module (LKM).

VaultFS has features that make it fully orthogonal to any literature proposal for data immutability—for example it does not require any hardware-level WORM support. It also offers an innovative support for Denial-of-Service (DOS) protection, which helps avoiding that un-trusted applications can render device blocks busy with the content of non-removable dummy files. It can therefore offer an additional option for data maintenance and accessibility on off-the-shelf devices, while still guaranteeing data integrity against ransomware and other data corruption attacks. Additionally, as we discussed it is highly configurable, hence giving an administrator the possibility to select the setup that better matches with his expectations—like for example the duration of the file protection lifetime, which can in turn lead to the re-usability of the blocks of the device for other contents.

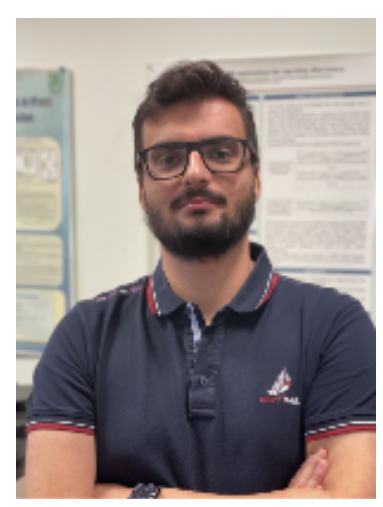

**Pasquale Caporaso** is a Security Researcher at CNIT and a PhD student at the University of Rome Tor Vergata, his research is focused on cyber-security and Operating Systems. He has accumulated extensive experience in the security field during his undergraduate years, worked as a Malware Analyst for the multinational Leonardo spa where he experienced first-hand the attack and defense of Linux Operating Systems, and, as a member of the Italian hacking team mhackeroni, has participated in numerous Italian and international competitions, including the finals of DEFCON and HackASat CTFs, some of largest hacking contests in the world.

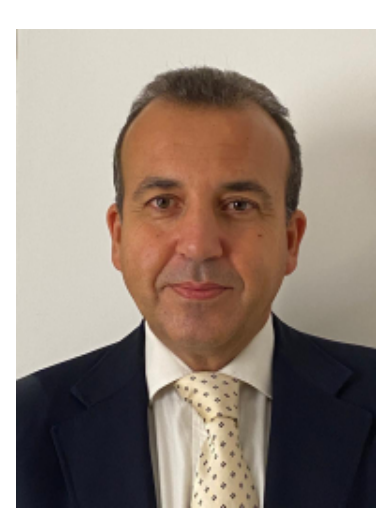

**Giuseppe Bianchi** is a Full Professor of Telecommunications and Network Security at the University of Roma Tor Vergata. His research interests encompass network security, network programmability, wireless networks, and in most generality any activity related to the design, analysis, monitoring and security of networked systems. He has documented his research through more than 300 publications in international venues, which have amassed over 20,000 citations according to Google Scholar. He has received numerous awards, including the ACM SIGMOBILE Test-of-Time Award in 2017 for his pioneering work on Wi-Fi system performance.

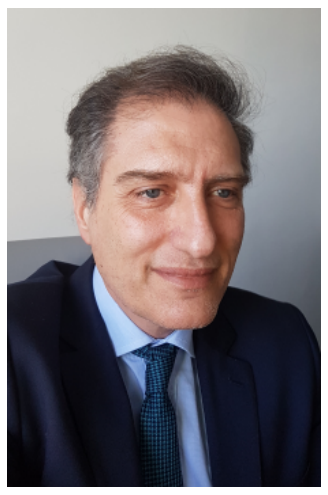

**Francesco Quaglia** received his MS in Electronic Engineering in 1995 and his PhD in Computer Engineering in 1999, both from Sapienza University of Rome, where he has worked as Assistant and Associate Professor from September 2000 till June 2017. Since then, he works as a Full Professor at the University of Rome Tor Vergata. His research interests include high performance and scalable computing, operating systems, and cyber-security. He has been the editor in chief of ACM TOMACS from 2019 to 2023, and has been involved as program and general chair in various IEEE and ACM conferences. He is (or has been) the principal investigator or the research unit leader in several projects at national and international (EU) levels.